\documentclass[graybox]{svmult}
\usepackage{mathptmx}       
\usepackage{helvet}         
\usepackage{courier}        
\usepackage{makeidx}         
\usepackage{graphicx}        
\usepackage{multicol}        
\usepackage[bottom]{footmisc}
\usepackage{amsmath,amssymb,amsfonts}        
\makeindex             
\usepackage{subcaption}
\usepackage{comment}

\title*{Thin filaments in Hele-Shaw cells}
\author{Nitay Ben-Shachar \and Michael C. Dallaston \and Scott W. McCue \thanks{
This report presents the results of a project undertaken by the first author at the Matrix Workshop \textit{Instabilities in Porous Media}, April 3-23, 2024, under the supervision of the other authors. 
}}
\institute{Nitay Ben-Schachar \at School of Mathematics \& Statistics, University of Melbourne \at \email{nbenshachar@student.unimelb.edu.au} 
\and Scott W. McCue \at School of Mathematical Sciences, Queensland University of Technology \at \email{scott.mccue@qut.edu.au}
\and Michael C. Dallaston \at School of Mathematical Sciences, Queensland University of Technology \at \email{michael.dallaston@qut.edu.au}
}
\authorrunning{N. Ben-Shachar \and M. C. Dallaston \and S. W. McCue}

\begin{document}

\maketitle

\section{Introduction}

Hele-Shaw cells consist of two parallel plates separated by a small gap.  The discrepancy between the in-plate and out-of-plate length-scales gives viscous-dominated Poiseuille flow between the two plates. Averaging over the depth between the plates, the macroscopic in-plate dynamics are governed by the incompressible Darcy law,
\begin{align}
    \mathbf{u} = -\frac{b^2}{12\mu} \nabla p , \quad \nabla \cdot \mathbf{u} = 0, \label{eq:GovEqs}
\end{align}
where $\mathbf{u}$ is the in-plate depth-averaged velocity, $p$ is the pressure, $b$ is the distance between the parallel plates and $\mu$ is the viscosity of the fluid~\cite{Paterson1981, Morrow2023, Morrow2021}. Thus, Hele-Shaw flows mimic the behaviour of two-dimensional fluid flows in porous media.

Hele-Shaw flows have direct industrial uses, such as the application of adhesives~\cite{Flaig2023}. Furthermore, they provide valuable insight to the governing dynamics of porous flows, which are typically in hard-to-access environments, such as the subsurface. Applications of subsurface, porous media flows include underground fluid mixing and CO$_2$ sequestration~\cite{Kampitsis2023}. 

\begin{figure}
    \centering
    \begin{subfigure}[l]{0.45\linewidth}
        \includegraphics[width=\linewidth]{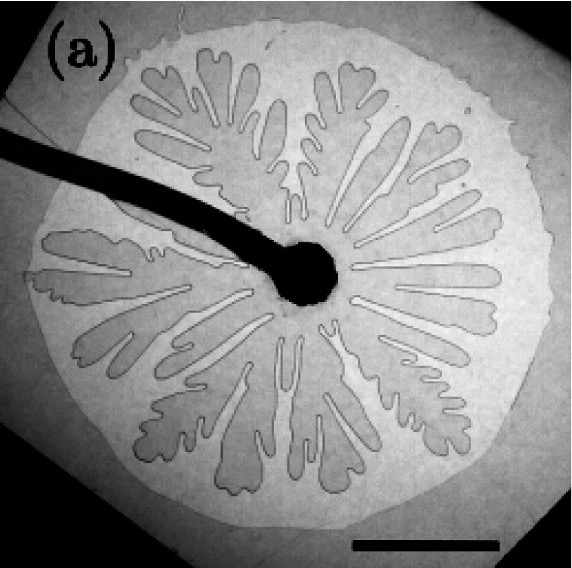}
        \caption{}
    \end{subfigure}
    \begin{subfigure}[l]{0.45\linewidth}
        \includegraphics[width=\linewidth]{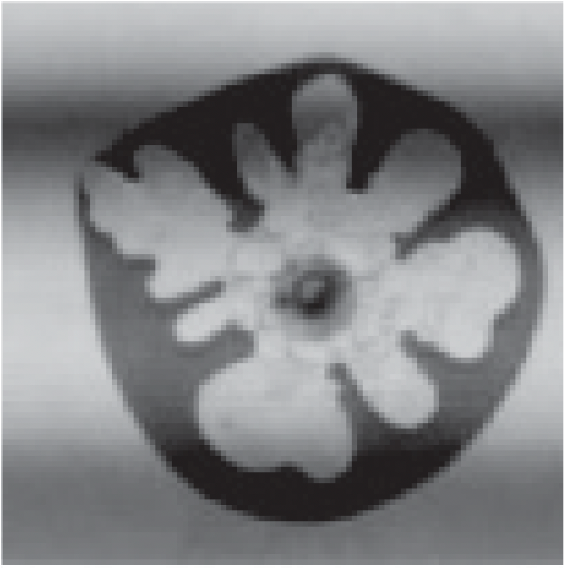}
        \caption{}
    \end{subfigure}
    \begin{subfigure}[l]{0.45\linewidth}
        \includegraphics[width=\linewidth]{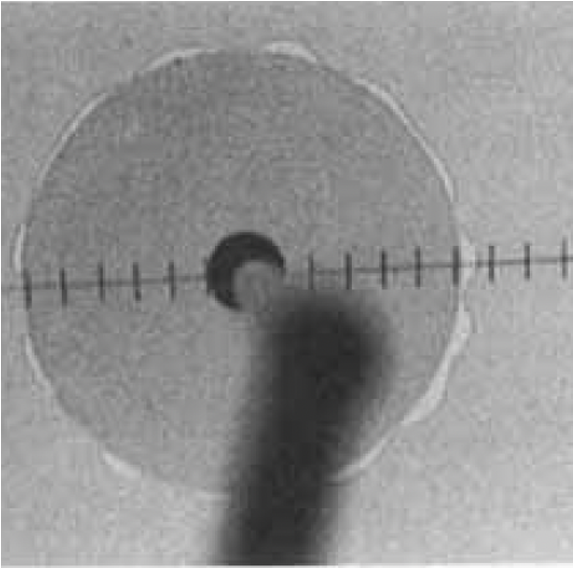}
        \caption{}
    \end{subfigure}
    \begin{subfigure}[l]{0.45\linewidth}
        \includegraphics[width=\linewidth]{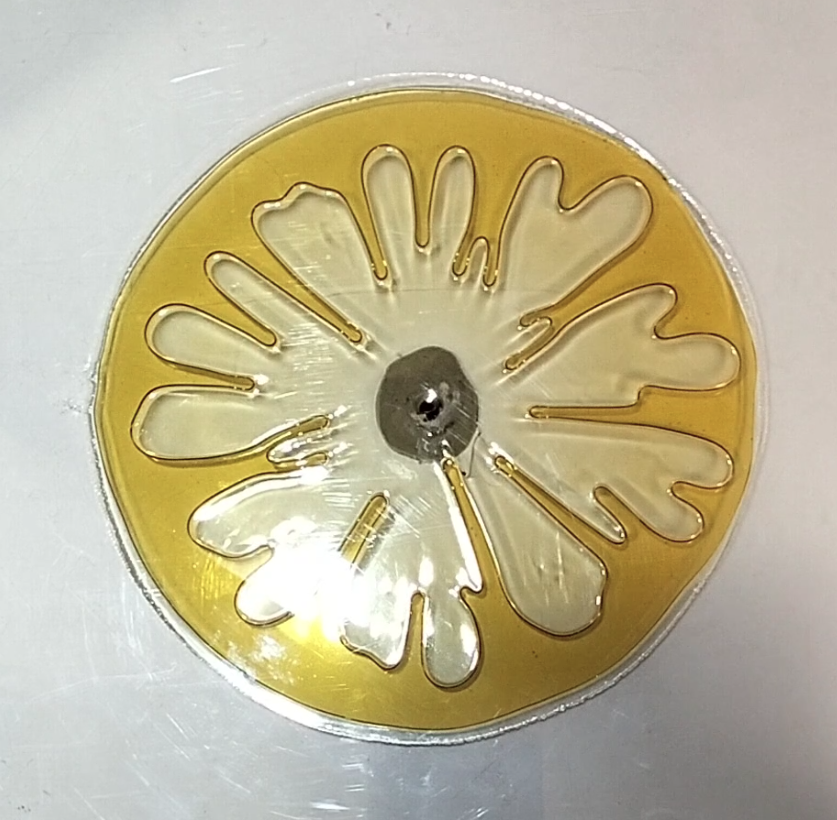}
        \caption{}
    \end{subfigure}
    \caption{Experiments of flows in a Hele-Shaw cell with multiple fluid interfaces; (a) Morrow {et al.}~\cite{Morrow2023}, (b) Ward and White~\cite{Ward2011}, (c) Cardoso and Woods~\cite{Cardoso1995}, (d) `home-made' experiments with Emily Chen.}
    \label{fig:experiments}
\end{figure}

\begin{figure}
    \centering
    \begin{subfigure}[t]{0.45\linewidth}
        \includegraphics[width=\linewidth]{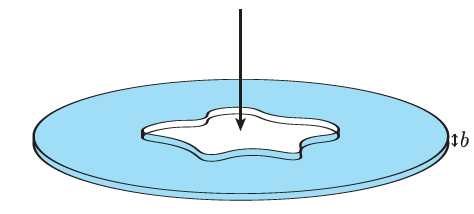}
        \caption{} \label{fig:inf_HS}
    \end{subfigure}
    \begin{subfigure}[t]{0.45\linewidth}
        \includegraphics[width=\linewidth]{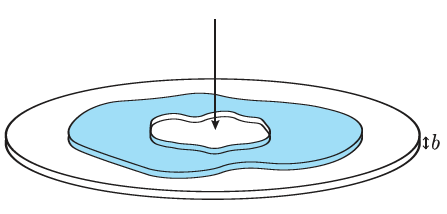}
        \caption{} \label{fig:finite_HS}
    \end{subfigure}
    \caption{Reprinted from Morrow {et al.}~\cite{Morrow2023}. (a) Injection of a bubble (white) into an infinite expanse of a viscous fluid (cyan) in a Hele-Shaw cell. (b) Fluid annulus (cyan) expanding or contracting due to a maintained pressure difference between the inside bubble and the ambient air.}
    \label{fig:HeleShawModel}
\end{figure}

Viscous fingering occurs in a Hele-Shaw cell when a viscous fluid is invaded by a less viscous fluid, as shown in Fig.~\ref{fig:experiments}. The interface between these fluids becomes unstable, and fingers form, grow and split. In contrast, when a viscous fluid invades a less viscous fluid, the interface is stable~\cite{Paterson1981}.
For the purposes of mathematical modelling, it is often assumed that the more viscous fluid is infinite in its extent, such as shown in Fig.~\ref{fig:inf_HS}, with a single interface dividing the two fluids. However, flows in real Hele-Shaw cells typically exhibit multiple fluid interfaces~\cite{Cardoso1995, Morrow2023}; see Fig.~\ref{fig:experiments}. Therefore, for real Hele-Shaw experiments a more realistic model has two interfaces, such as shown in Fig.~\ref{fig:finite_HS}. Morrow {et al.}~\cite{Morrow2023} demonstrated, by studying the injection of air into a finite amount of water in a Hele-Shaw cell, that both inner and outer interfaces can become unstable. When the distance between the inner and outer interfaces is comparable to the Hele-Shaw cell gap, the depth-averaged model fails and the two interfaces may merge (i.e., the inner air bubble bursts the fluid annulus).

Recently, Dallaston {et al.}~\cite{Dallaston2024} developed general theory for the evolution of thin liquid filaments in Hele-Shaw cells, whereby the two interfaces are very close together. Here, the thickness of the liquid filament is much smaller than the radius of curvature of either interface, but significantly thicker than the Hele-Shaw cell gap such that the depth-averaged model applies.
\begin{figure}
    \centering
    \includegraphics[width=0.5\linewidth]{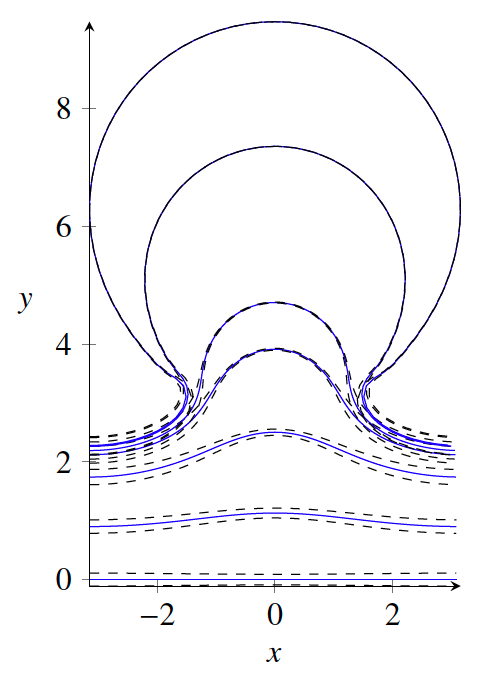}
    \caption{Reprinted from Dallaston {et al.}~\cite{Dallaston2024}. Evolution of a perturbed straight filament, driven by a constant pressure gradient. An instability results in the interface stretching and thinning, until ultimately is appears that the filament approaches a circular shape. The blue line marks the centre of the filament, with the dashed black lines indicating the filament boundaries.}
    \label{fig:Dallaston}
\end{figure}
It was found that, in the absence of a bursting criteria (i.e., for an infinitesimally small Hele-Shaw cell gap), perturbations to straight, thin filaments grow into circular-like shapes, as can be seen in Fig.~\ref{fig:Dallaston}. However, these circular filament shapes have only been observed numerically, and their dynamics and stability are currently unknown analytically. Further, the shrinking filament thickness induces numerical instabilities for later times, making it difficult to glean the driving mechanism for the dynamics of these filament structures.

In Sect.~\ref{sec:model} we summarise the general Hele-Shaw model~\cite{Morrow2023} and the filament model from Dallaston {et al.}~\cite{Dallaston2024}. These are used to determine axisymmetric annular fluid solutions and to study their stability in Sect.~\ref{sec:annulus}. In Sect.~\ref{sec:PC} we derive a growing and translating solution to the filament model, describing the circle-like structures observed in numerical solutions. Conclusions and future work are discussed in Sect.~\ref{sec:conclusion}.

\section{Model outline}\label{sec:model}

Consider the flow of a finite volume of fluid in a Hele-Shaw cell. We non-dimensionalise all variables by selecting the following scales:
\begin{align}
    x_s = \frac{l}{2\pi}, \quad p_s = \frac{12 \mu x_s}{b^2 t_s},
\end{align}
where $x_s$ is the spatial scale, $p_s$ is the pressure scale, $l$ is some macroscopic length of the fluid shape that will be chosen later, and $t_s$ is an arbitrary timescale.

\subsection{Full Hele-Shaw model}

We consider the dynamics of an annular section of liquid, confined by air (i.e.~inviscid fluid) on both the interior and exterior boundaries, see Fig.~\ref{fig:HeleShawModel}~\cite{Dallaston2024,Morrow2023}. Scaling the governing equations,~\eqref{eq:GovEqs}, gives
\begin{subequations}
\begin{align}
    \nabla^2 p & = 0 , && \mathbf{x}\in \Omega , \label{eq:HS_eqn_1}\\
    v_n & = -\nabla p\cdot \mathbf{\hat{n}} , && \mathbf{x} \in \partial \Omega , \label{eq:HS_eqn_2}\\
    p & = \Delta P - \sigma \kappa , && \mathbf{x} \in \partial \Omega_\text{in} , \label{eq:HS_eqn_3}\\
    p & = \sigma \kappa , && \mathbf{x} \in \partial \Omega_\text{out} , \label{eq:HS_eqn_4}
\end{align}
\end{subequations}
where we have invoked appropriate pressure-jump conditions at the inner and outer interfaces, $\Delta P$ is the scaled driving pressure difference across the annulus, $\sigma = \hat{\gamma}/(p_s x_s)$ is the scaled surface tension, $v_n$ is the boundary-normal velocity, $\mathbf{\hat{n}}$ is the unit boundary normal, and $\kappa$ is the curvature of the boundary. The inner and outer boundaries are $\partial \Omega_\text{in}$ and $\partial \Omega_\text{out}$, respectively, while the fluid domain is $\Omega$. With an appropriate choice for the timescale, $t_s$, either of $\Delta P$ or $\sigma$ (if nonzero) could be scaled to unity. However, we retain both parameters explicitly to aid in interpreting the terms that arise in our analysis.

Other models of multi-layer radial Hele-Shaw flows have been proposed and studied, for example in~\cite{Anjos2020,Crowdy2004,Dallaston2012,Gin2021,Gin2021b,Zhao2020}. Gin and Daripa~\cite{Gin2021b,Gin2021} studied the linear stability of radial multi-layer Hele-Shaw cells, while Anjos and Li~\cite{Anjos2020} used weakly-nonlinear stability analysis to predict the morphology of the viscous fingers that form. Zero-surface-tension models were considered by Crowdy and Tanveer~\cite{Crowdy2004}, and Dallaston and McCue~\cite{Dallaston2012}, where exacts solutions were found through use of complex variable techniques. Zhao {et al.}~\cite{Zhao2020} considered a radial three-layer Hele-Shaw system, with the outer fluid more viscous than the middle fluid, which in turn is more viscous than the inner fluid. The model described by Morrow {et al.}~\cite{Morrow2023} and presented here, where a viscous fluid is confined on both inner and outer boundaries by an inviscid fluid, complements the work of Zhao {et al.}~\cite{Zhao2020}.

\subsection{Filament model}

The filament model applies when the distance between the outer and inner surfaces (the film thickness, $h$) is small compared to the radius of curvature of either interface. The film thickness is assumed to be significantly larger than the plate separation, $b$, such that the depth-averaged Hele-Shaw model is valid. In this limit,~\eqref{eq:HS_eqn_1}--\eqref{eq:HS_eqn_4} can be approximated with a lubrication-style method, giving explicit equations for the evolution of the filament centreline and the thickness of the filament along the centreline, summarised below in Sect.~\ref{sec:FFM}. Resolving the governing equations to first-order in the lubrication parameter yields the so-called `full filament model'. However, the key model features can be captured by neglecting many of the derived terms, resulting in the reduced `regularised leading-order filament model'. As we will see in Sect.~\ref{sec:LSA_filament}, the regularised leading-order filament model qualitatively agrees with the full Hele-Shaw calculations,~\eqref{eq:HS_eqn_1}--\eqref{eq:HS_eqn_4}, and allows for physical interpretation of the model predictions.

\subsubsection{Full filament model}\label{sec:FFM}

The full filament model is derived by Dallaston {et al.}~\cite{Dallaston2024} as
\begin{subequations}
\begin{align}
    v_n & = v_0 +\left(\frac{2h^2\kappa^2 v_0-h^2v_{0ss}}{12}-\frac{hh_sv_{0s}}{4}+\frac{2\sigma\kappa_2}{h}\right) , \label{eq:FFM_vnEqn}\\
    \frac{Dh}{Dt} & = \kappa h v_n-\frac{\partial}{\partial s}\left[-\frac{h^3\kappa_sv_0}{12}-\frac{h^2h_s\kappa v_0}{4}+\sigma h\kappa_{1s}\right] , \label{eq:FFM_hEqn}
\end{align}
\end{subequations}
with
\begin{subequations}
\begin{align}
    v_0 & = \frac{\Delta P+2\sigma \kappa}{h}, \\
    \kappa_1 & = \frac{h\kappa^2}{2}+\frac{h_{ss}}{2}, \\
    \kappa_2 & = \frac{h^2\kappa^3}{4}+\frac{h h_s\kappa_s}{4}+\frac{h h_{ss}\kappa}{2}+\frac{h_s^2\kappa}{8} ,
\end{align}
\end{subequations}
where $\kappa$ is the curvature of the centreline, $s$ is the arclength variable parameterising the centreline, and $D/Dt$ is a time derivative along the normal direction to the filament centreline. The first terms on the right hand side of~\eqref{eq:FFM_vnEqn} and~\eqref{eq:FFM_hEqn} are the leading-order solutions in the lubrication approximation, while the remainder of the terms arise from continuing the expansion to first order in the lubrication parameter.

\subsubsection{Regularised leading-order filament model} \label{sec:RLOFM}

The regularised leading-order model is obtained by neglecting all terms of order the lubrication parameter in~\eqref{eq:FFM_vnEqn} and~\eqref{eq:FFM_hEqn} except for the regularising term, $\sigma (h h_{sss})_s/2$, giving
\begin{subequations}
\begin{align}
    v_n = & v_0 , \label{eq:RLOFM_eqn1}\\
    \frac{Dh}{Dt} = & \kappa h v_n - \frac{\sigma}{2}\frac{\partial}{\partial s} \left(hh_{sss}\right) . \label{eq:RLOFM_eqn2}
\end{align}
\end{subequations}
The necessity of retaining the regularising term will be discussed further in Sect.~\ref{sec:LSA_filament}. When surface tension is neglected from~\eqref{eq:RLOFM_eqn1} and~\eqref{eq:RLOFM_eqn2}, the zero-surface-tension filament model of Farmer and Howison~\cite{Farmer2006} is obtained.

\section{Linear stability analysis of the annulus} \label{sec:annulus}

Here we consider the linear stability of an annulus of fluid in a Hele-Shaw cell driven by a constant pressure difference between the inside and outside interfaces. We will begin by considering an annulus of arbitrary thickness, followed by detailed analysis of the thin annulus where the filament model applies.

\subsection{Full Hele-Shaw model} \label{sec:2interfaces}

Parameterising the inner and outer fluid interfaces with a polar angle, $\theta$, the curvature of each boundary is
\begin{align}
    \kappa_i = \frac{2(R_{i}')^2-R_{i}R_{i}''+R_i^2}{\left(R_i^2+(R_i')^2\right)^{3/2}},
\end{align}
where $R_{I}(t,\theta)$ and $R_{O}(t,\theta)$ are the radial locations of the inner and outer boundaries, respectively, $i\in\{I,O\}$ and the prime symbol, $'$, denotes differentiation with respect to $\theta$. The boundary unit normal, pointing into the fluid domain, is
\begin{align}
    \mathbf{\hat{n}} = \pm\frac{\hat{r}-\hat{\theta}(R_i'/R_i)}{\sqrt{1+\left((R_i'/R_i)\right)^2}},
\end{align}
where the positive and negative signs are taken for the inner and outer boundaries, respectively.

\subsubsection{Base state - axisymmetric solution}\label{sec:HS_base}

The base, axisymmetric solution to~\eqref{eq:HS_eqn_1}--\eqref{eq:HS_eqn_4}, denoted with a subscript `$b$', satisfies
\begin{subequations}
\begin{align}
    \frac{\D^2 p_b}{\D r^2} + \frac{1}{r}\frac{\D p_b}{\D r} = & 0 , \label{eq:HS_baseEqn1}\\
    p_b(R_{bI}) = & \Delta P- \frac{\sigma}{R_{bI}} , \label{eq:HS_baseEqn2}\\
    p_b(R_{bO}) = & \frac{\sigma}{R_{bO}} , \label{eq:HS_baseEqn3}\\
    \frac{\D R_{bI}}{\D t} = & -\left.\frac{\D p_b}{\D r}\right|_{r=R_{bI}} , \label{eq:HS_baseEqn4} \\
    \frac{\D R_{bO}}{\D t} = & -\left.\frac{\D p_b}{\D r}\right|_{r=R_{bO}} . \label{eq:HS_baseEqn5}
\end{align}
\end{subequations}
Solving~\eqref{eq:HS_baseEqn1}--\eqref{eq:HS_baseEqn3} furnishes the pressure within the fluid,
\begin{align}
    p = \frac{1}{R_{bO}}+\left[ \frac{\Delta P - \sigma\left(R_{bO}^{-1}+R_{bI}^{-1}\right)}{\log \left(R_{bI}/R_{bO}\right)}\right] \log \left(\frac{r}{R_{bO}}\right) . \label{eq:HSb_PSoln}
\end{align}
Substituting~\eqref{eq:HSb_PSoln} into~\eqref{eq:HS_baseEqn4} and~\eqref{eq:HS_baseEqn5} gives the evolution equations for the inner and outer interfaces~\cite{Morrow2023},
\begin{subequations}
\begin{align}
    \frac{\D R_{bI}}{\D t} & = \frac{\Delta P - \sigma\left(R_{bO}^{-1}+R_{bI}^{-1}\right)}{R_{bI} \log \left(R_{bO}/R_{bI}\right)} , \label{eq:HS_sbIt}\\
    \frac{\D R_{bO}}{\D t} & = \frac{\Delta P - \sigma\left(R_{bO}^{-1}+R_{bI}^{-1}\right)}{R_{bO} \log \left(R_{bO}/R_{bI}\right)} . \label{eq:HS_sbOt}
\end{align}
\end{subequations}

Equations~\eqref{eq:HS_sbIt} and~\eqref{eq:HS_sbOt} reveal a competition between the applied pressure difference between the inner and outer regions of the annulus (pushing the interfaces away from the origin), and the surface tension (pulling the interfaces toward the origin). The axisymmetric solution grows if $\Delta P > \sigma (R_{bI}^{-1}+R_{bO}^{-1})$, and shrinks if the inequality is reversed.

\subsubsection{Linear perturbation}

We now turn to the linear stability of the base solution derived in Sect.~\eqref{sec:HS_base}. The solution is written as a sum of the base solution and a small perturbation (subscript `$p$'),
\begin{subequations}
\begin{align}
    p = p_b(r,t) + \epsilon p_p(r,\theta,t), \label{eq:HS_pertExpan1}\\
    R_{O} = R_{bO}(t) + \epsilon R_{pO}(t)\cos(n\theta) , \label{eq:HS_pertExpan2}\\
    R_{I} = R_{bI}(t) + \epsilon R_{pI}(t) \cos(n\theta) , \label{eq:HS_pertExpan3}
\end{align}
\end{subequations}
and $\epsilon$ is the small perturbation amplitude. Substituting~\eqref{eq:HS_pertExpan1}--\eqref{eq:HS_pertExpan3} into~\eqref{eq:HS_eqn_1}--\eqref{eq:HS_eqn_4}, expanding for $\epsilon \ll 1$ and collecting powers of $\epsilon$, the governing equations and boundary conditions at $O(\epsilon)$ are
\begin{subequations}
\begin{align}
    \frac{\partial^2 p_p}{\partial r^2} + \frac{1}{r} \frac{\partial p_p}{\partial r} + \frac{1}{r^2}\frac{\partial^2 p_p}{\partial^2\theta} & = 0 , \\
    p_p(r_I, \theta, t) & =-R_{pI}\frac{\D R_{bI}}{\D t}+\sigma\frac{R_{pI}(1-n^2)}{R_{bI}^2} ,\\
    p_p(r_O, \theta, t) & =-R_{pO}\frac{\D R_{bO}}{\D t}-\sigma\frac{R_{pO}(1-n^2)}{R_{bO}^2} , \\
    \frac{\D R_{pI}}{\D t} & =-\left.\frac{\partial p_p}{\partial r}\right|_{r=R_{bI}}-R_{pI} \left.\frac{\D^2 p_b}{\D r^2} \right|_{r=R_{bI}} , \\
    \frac{\D R_{pO}}{\D t} & =-\left.\frac{\partial p_p}{\partial r}\right|_{r=R_{bO}}-R_{pO} \left.\frac{\D^2 p_b}{\D r^2} \right|_{r=R_{bO}} .
\end{align}
\end{subequations}
Solving these, we obtain the evolution equations for $R_{pI}$ and $R_{pO}$,
\begin{subequations}
\begin{align}
    \frac{\D R_{pI}}{\D t} = & \left[\frac{1}{R_{bI}}\left(\frac{n(R_{bI}^{2n}+R_{bO}^{2n})}{R_{bO}^{2n}-R_{bI}^{2n}}-1\right)\frac{\D R_{bI}}{\D t}-\sigma\frac{n (n^2-1)(R_{bI}^{2n}+R_{bO}^{2n})}{R_{bI}^3(R_{bO}^{2n}-R_{bI}^{2n})}\right] R_{pI} \nonumber \\& - \frac{2nR_{bI}^nR_{bO}^n}{R_{bO}^{2n}-R_{bI}^{2n}}\left[\frac{1}{R_{bO}}\frac{\D R_{bI}}{\D t}+\sigma\frac{n^2-1}{R_{bI}R_{bO}^2}\right] R_{pO} , \label{eq:spIt} \\
    \frac{\D R_{pO}}{\D t} & = \left[-\frac{1}{R_{bO}}\left(\frac{n(R_{bI}^{2n}+R_{bO}^{2n})}{R_{bO}^{2n}-R_{bI}^{2n}}+1\right)\frac{\D R_{bO}}{\D t}-\sigma\frac{n (n^2-1)(R_{bI}^{2n}+R_{bO}^{2n})}{R_{bO}^3(R_{bO}^{2n}-R_{bI}^{2n})}\right] R_{pO} \nonumber \\& + \frac{2nR_{bI}^nR_{bO}^n}{R_{bO}^{2n}-R_{bI}^{2n}}\left[\frac{1}{R_{bI}}\frac{\D R_{bO}}{\D t}-\sigma\frac{n^2-1}{R_{bO}R_{bI}^2}\right] R_{pI} . \label{eq:spOt}
\end{align}
\end{subequations}
These are identical to the results reported by Morrow {et al.}~\cite{Morrow2023}. While difficult to interpret,~\eqref{eq:spIt} and~\eqref{eq:spOt} will be used as a benchmark for the filament model, which will be analysed next.

\subsection{Linear stability of the filament model} \label{sec:filament_axisymmetric}

We now turn to the regularised leading-order filament model derived by Dallaston {et al.}~\cite{Dallaston2024}, summarised in Sect.~\ref{sec:RLOFM}. 

\subsubsection{Base state - axisymmetric solution}

As in Sect.~\ref{sec:2interfaces}, the base state, denoted subscript $b$, is found by searching for axisymmetric solutions of the governing equations \eqref{eq:RLOFM_eqn1} and~\eqref{eq:RLOFM_eqn2}, i.e., where the centreline radius and filament thickness are arc-length independent. The axisymmetric solution satisfies
\begin{subequations}
\begin{align}
    \frac{\D R_b}{\D t} = \frac{\Delta P-2\sigma/R_b}{h_b}, \label{eq:RLOFM_baseEqn1}\\
    \frac{\D h_b}{\D t} = -\frac{h_b}{R_b} \frac{\D R_b}{
    \D t} , \label{eq:RLOFM_baseEqn2}
\end{align}
\end{subequations}
where $R_b$ is the centreline radius of the axisymmetric (base) solution. Taking the ratio of~\eqref{eq:RLOFM_baseEqn1} and~\eqref{eq:RLOFM_baseEqn2} and integrating with respect to time, we find the standard conservation of volume condition:
\begin{align}
    h_b(t) R_b(t) = h_b(0)R_b(0) = \frac{A}{2\pi}, \label{eq:RLOFM_CoM}
\end{align}
where $A$ is the dimensionless surface area of the filament. Substituting~\eqref{eq:RLOFM_CoM} into~\eqref{eq:RLOFM_baseEqn1} and~\eqref{eq:RLOFM_baseEqn2} and integrating furnishes the centreline radius and film thickness solutions, respectively:
\begin{subequations}
\begin{align}
    R_b(t) = & \left(R_b(0)-\frac{2\sigma}{\Delta P}\right)\exp \left(\frac{2\pi\Delta P t}{A}\right)+\frac{2\sigma}{\Delta P} , \\
    h_b(t) = & \frac{h_b(0) R_b(0)}{\left(R_b(0)-2\sigma/\Delta P\right)\exp\left(2\pi\Delta P t/A\right)+2\sigma/\Delta P} .
\end{align}
\end{subequations}

As observed in the full filament model (Sect.~\ref{sec:2interfaces}), the axisymmetric solution is characterised by a competition between the applied pressure difference and the surface tension. The filament radius grows if the initial radius exceeds the critical radius $R_c$:
\begin{align}
    R>R_c= \frac{2\sigma}{\Delta P}, \label{eq:sc_defn}
\end{align}
and shrinks if the inequality is reversed. The growing filament radius increases exponentially, due to the shrinking thickness of the filament. In contrast, a filament whose radius shrinks becomes thicker exponentially. As such, the filament model will eventually become inapplicable for a shrinking filament, when its thickness becomes comparable to its radius.

\subsubsection{Linear perturbation} \label{sec:LSA_filament}

\begin{figure}
    \centering
    \begin{subfigure}[c]{\linewidth}
        \includegraphics[width=\linewidth]{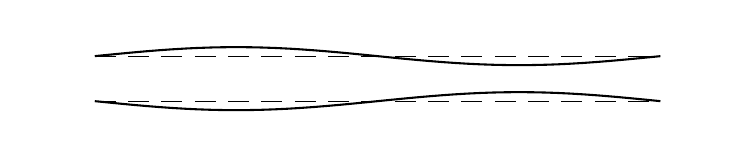}
        \caption{}
    \end{subfigure}
    \begin{subfigure}[c]{\linewidth}
        \includegraphics[width=\linewidth]{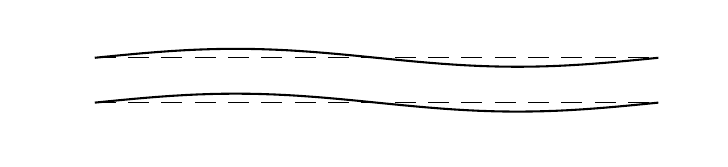}
        \caption{}
    \end{subfigure}
    \caption{Inner and outer interfaces of a filament perturbed from their equilibrium positions (dashed lines) for: (a) $h_p=1, R_p=0$, (b) $h_p=0, R_p=1$.}
    \label{fig:pert_basis}
\end{figure}

Consider a linear perturbation of the annular filament solution derived in the previous section. We write the radius and thickness of the filament as
\begin{subequations}
\begin{align}
    R(t,\theta) = & R_b(t) + \epsilon R_p(t)\cos(n\theta) , \label{eq:RLOFM_pertExpan1} \\
    h(t,\theta) = & h_b(t) + \epsilon h_p(t) \cos(n\theta) .\label{eq:RLOFM_pertExpan2}
\end{align}
\end{subequations}
While we have assumed a restricted form for the linear perturbations, any desired filament perturbation can be obtained with an appropriate superposition of the perturbations,~\eqref{eq:RLOFM_pertExpan1} and~\eqref{eq:RLOFM_pertExpan2}, with varying $n$. The chosen basis of the perturbations is shown in Fig.~\ref{fig:pert_basis}.

Substituting~\eqref{eq:RLOFM_pertExpan1} and~\eqref{eq:RLOFM_pertExpan2} into~\eqref{eq:RLOFM_eqn1} and~\eqref{eq:RLOFM_eqn2} and expanding to linear order for $\epsilon \ll 1$, we obtain the governing equations for the perturbations of the film thickness and radius, respectively:
\begin{subequations}
\begin{align}
    \frac{\D h_p}{\D t} = & -R_p \frac{n^2-1}{R_b^2} \left(h_b \frac{\D R_b}{\D t}-\frac{2\sigma}{R_b}\right)-\frac{\sigma h_b n^4}{2R_b^4} h_p , \label{eq:RLOFM_pertEq1} \\
    \frac{\D R_p}{\D t} & = -\frac{4\pi\sigma (n^2-1)}{A R_b} R_p - \frac{\D  R_b}{\D t}\frac{1}{h_b} h_p . \label{eq:RLOFM_pertEq2}
\end{align}
\end{subequations}
Using a `frozen time analysis', the eigenvalues of the system described by~\eqref{eq:RLOFM_pertEq1} and~\eqref{eq:RLOFM_pertEq2} give the local growth rate of the perturbations $h_p$ and $R_p$:
\begin{align}
    \lambda_\pm = & -\left(\frac{A n^4 \sigma}{8\pi R_b^5}+\frac{2\pi\left(n^2-1\right) \sigma}{A R_b}\right) \pm \frac{1}{2}\Bigg[ \left(\frac{A n^4 \sigma}{4\pi R_b^5}+\frac{4\pi \left(n^2-1\right) \sigma}{A R_b}\right)^2\nonumber\\ &-\frac{4(n^2-1)}{R_b^2} \left(\frac{ n^4 \sigma^2}{R_b^4}-\frac{4\pi^2(R_b \Delta P-4 \sigma) (R_b \Delta P-2 \sigma)}{A^2}\right)\Bigg]^{1/2} . \label{eq:lambdaSol}
\end{align}
In the large radius limit, $R_b\to \infty$,~\eqref{eq:lambdaSol} gives the limiting eigenvalues
\begin{align}
    \lambda_\pm = \pm \frac{\Delta P}{A}\sqrt{n^2-1} + O\left(\frac{1}{R_b}\right).
\end{align}
From~\eqref{eq:lambdaSol} we find a critical mode number for stability,
\begin{align}
    n_c = \frac{R_b}{\sqrt{A \sigma}}\left((R_b\Delta P-4\sigma)(R_b \Delta P-2\sigma)\right)^{1/4} . \label{eq:nc}
\end{align}
When $n_c>1$, all wavenumbers satisfying $1<n<n_c$ are unstable, while those satisfying $n>n_c$ are stable. Hence, all wavenumbers are stable when
\begin{align}
    R_b<\frac{4\sigma}{\Delta P}=2R_c. \label{eq:stable_sb}
\end{align}
In the absence of the regularising term in the leading-order filament model (i.e., neglecting all $n^4$ terms in~\eqref{eq:lambdaSol}), the (non-regularised) leading-order filament model growth rates are
\begin{align}
    \lambda_\pm^\text{LO} = -\frac{2\pi \sigma(n^2-1)\pm 2\pi \sqrt{n^2-1} \sqrt{(R_b \Delta P-4\sigma)(R_b \Delta P-2\sigma) + \left(n^2-1\right) \sigma^2}}{A R_b}. \label{eq:lambdaLO}
\end{align}

\begin{figure}
    \centering
    \begin{subfigure}[c]{0.49\linewidth}
        \includegraphics[width=\linewidth]{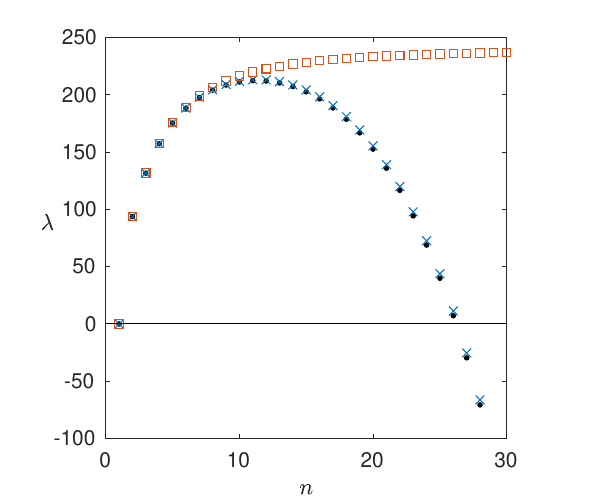}
        \caption{}
    \end{subfigure}
    \begin{subfigure}[c]{0.49\linewidth}
        \includegraphics[width=\linewidth]{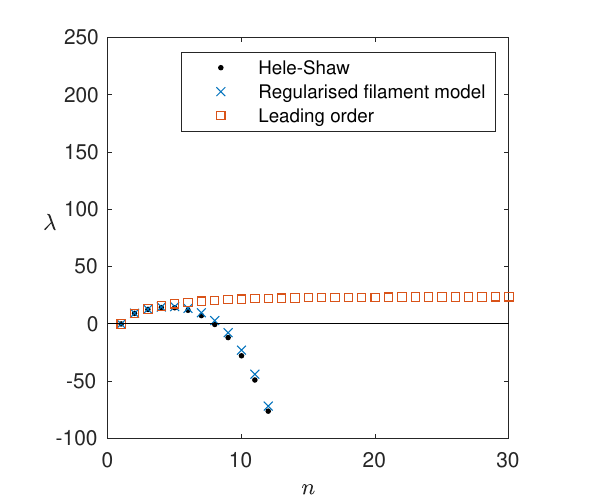}
        \caption{}
    \end{subfigure}
    \caption{Largest growth rate $\lambda$ of perturbations from the annulus of given mode number $n$ as predicted by the Hele-Shaw two-interface model, \eqref{eq:spIt} and~\eqref{eq:spOt}; regularised leading-order filament model,~\eqref{eq:lambdaSol}; and the (non-regularised) leading-order filament model,~\eqref{eq:lambdaLO}. The growth rates are shown for $\Delta P=1, \sigma=0.1, R_b=1$ and (a) $h_b=0.01$, (b) $h_b=0.1$.}
    \label{fig:lambda}
\end{figure}

The growth rates predicted by the full Hele-Shaw model (calculated by numerically diagonalising~\eqref{eq:spIt} and~\eqref{eq:spOt}), regularised leading-order filament model~\eqref{eq:lambdaSol}, and (non-regularised) leading-order filament model~\eqref{eq:lambdaLO}, are shown in Fig.~\ref{fig:lambda} for selected parameter values. The regularised leading-order filament model closely predicts the full Hele-Shaw model for the given parameter values. For small mode numbers $n$, these exhibit positive growth rates, indicating unstable modes. For mode numbers greater than a critical `cut-off' mode number, given by~\eqref{eq:nc} for the regularised leading-order model, the growth rates are negative, indicating stable modes. In contrast, the leading-order model does not have a cutoff mode number; all modes are either stable (if $R_b<4\sigma/\Delta P$) or unstable (if $R_b>4\sigma/\Delta P$).

\begin{figure}
    \centering
    \begin{subfigure}[c]{0.45\linewidth}
        \includegraphics[width=\linewidth]{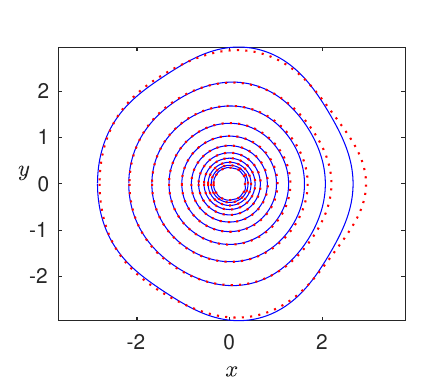}
        \caption{} \label{fig:integrated_linStab_a}
    \end{subfigure}
    \begin{subfigure}[c]{0.45\linewidth}
        \includegraphics[width=\linewidth]{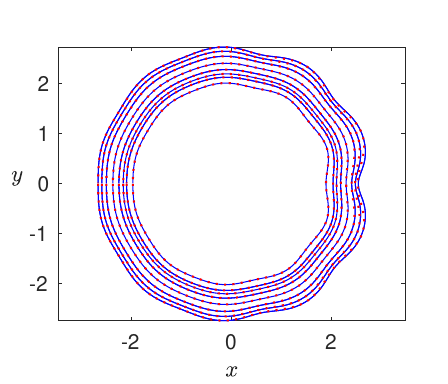}
        \caption{}\label{fig:integrated_linStab_b}
    \end{subfigure}
    \caption{Filament centrelines after a random perturbation of magnitude $1$\% to the centreline, shown at evenly spaced time intervals, as calculated using the full filament model (solid blue lines,~\eqref{eq:FFM_vnEqn} and~\eqref{eq:FFM_hEqn}) and integrating the linear stability analysis (dotted red lines,~\eqref{eq:RLOFM_pertEq1} and~\eqref{eq:RLOFM_pertEq2}). Parameter values are $\Delta P=1, \sigma=0.1, h_b(0)=0.05$, and (a) $R_b(0)=0.35$, (b) $R_b(0)=2$.}
    \label{fig:integrated_linStab}
\end{figure}

Next, we utilise the linear stability analysis to predict the perturbed filament dynamics. Initialising an annular filament with a perturbed centreline but constant thickness, composed of a random combination of the first 10 modes, we then numerically integrate~\eqref{eq:RLOFM_pertEq1} and~\eqref{eq:RLOFM_pertEq2} for the mode amplitudes. In Fig.~\ref{fig:integrated_linStab} we compare the results of this analysis with a numerical solution to the full filament model,~\eqref{eq:FFM_vnEqn} and~\eqref{eq:FFM_hEqn}~\cite{Dallaston2024}.

At early times the integrated linear stability analysis matches the full filament model calculations. For the smaller annulus shown in Fig.~\ref{fig:integrated_linStab}, the mode amplitudes initially decay, before the amplitude of various modes grows when a critical radius is reached. In contrast, for the larger annulus in Fig.~\ref{fig:integrated_linStab}, higher mode numbers dominate the centreline behaviour, as predicted by the linear stability analysis. The discrepancy between the integrated linear stability analysis and the full filament model increases with time, due to departure from the linear regime. However, the onset of fingering is well predicted by the linear theory, with the final centreline shape dominated by small ($n\lesssim 5$) and large ($n\gtrsim 5$) mode numbers, for Figs.~\ref{fig:integrated_linStab_a} and~\ref{fig:integrated_linStab_b}, respectively.

\section{Pinned circles} \label{sec:PC}

\begin{figure}
    \centering
    \begin{subfigure}[b]{0.45\linewidth}
        \includegraphics[width=\linewidth]{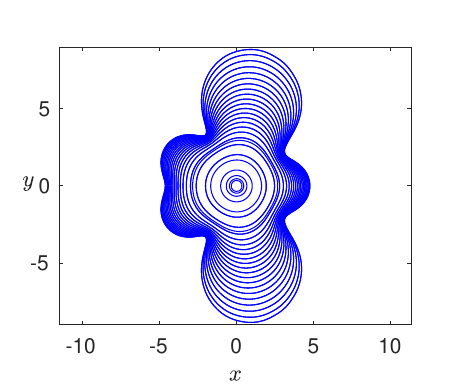}
        \caption{} \label{fig:nonlinearGrowth}
    \end{subfigure}
    \begin{subfigure}[b]{0.45\linewidth}
        \includegraphics[width=\linewidth]{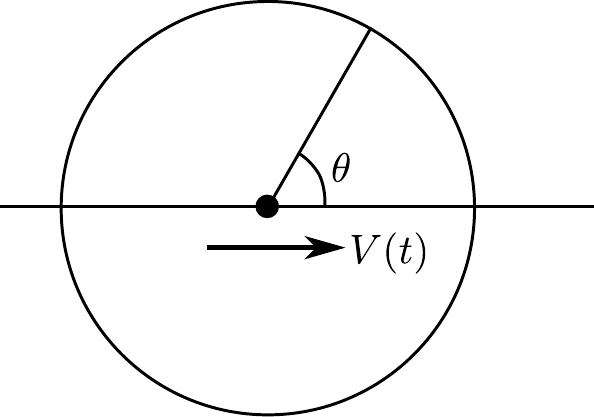}
        \vspace{0.25em}\caption{} \label{fig:PC_sketch}
    \end{subfigure}
    \caption{(a) Centreline of the filament calculated by the full filament model,~\eqref{eq:FFM_vnEqn} and~\eqref{eq:FFM_hEqn}~\cite{Dallaston2024}, for $\Delta P=1, \sigma=0.1, h_b(0)=0.05$ and $R_b(0)=0.35$. The shown centrelines are unevenly spaced in time to show the expanding structure of the centreline. (b) Sketch of a `pinned circle' whose centre translates with speed $V(t)$.}
\end{figure}

As described in the previous section, an axisymmetric circular filament becomes unstable for sufficiently large radii, so that modes of perturbation grow for small time. However, for longer times the non-linear growth of these perturbations appears circular away from their attachment points to the remainder of the filament; see Fig.~\ref{fig:nonlinearGrowth}. These dynamics share qualitative and quantitative features with the circular-like growth observed for perturbed flat filaments~\cite{Dallaston2024} (see Fig.~\ref{fig:Dallaston}): asymmetric radial growth, asymptoting to a circular geometry with a translating centre and increasing radius. These filament sections, henceforth referred to as `pinned circles', appear stable for radii significantly larger than the critical stable radius of the axisymmetric filament. We thus search for non-axisymmetric solutions of the filament model that could describe these pinned circles.

\subsection{Approximate solution}

The asymmetric growth of the pinned circles implies, from~\eqref{eq:FFM_vnEqn}, a non-constant filament thickness along the circle. We write~\eqref{eq:FFM_vnEqn} and~\eqref{eq:FFM_hEqn} in polar coordinates whose origin translates with the pinned circle at velocity $V(t)\hat{x}$ (see Fig.~\ref{fig:PC_sketch}). Searching for circular solutions in the large radius limit, i.e., $R\gg 1$, in which the zero-surface-tension model is appropriate, the governing equations for the thickness, $h(\theta, t)$ and radius, $R(t)$, are
\begin{subequations}
\begin{align}
    \frac{\partial h}{\partial t} = & -\frac{\Delta P}{R}-\frac{\partial h}{\partial \theta}\frac{V(t)\sin\theta}{R} , \\
    \frac{\D R}{\D t} = & \frac{\Delta P}{h}-V(t) \cos\theta .
\end{align}
\end{subequations}
These are solved to give
\begin{subequations}
\begin{align}
    V(t) = & \frac{\D R}{\D t}, \label{eq:PCsol1}\\
    h(\theta, t) = & \frac{\Delta P (t_0-t)^2}{R(0)t_0(1+\cos\theta)}, \label{eq:PCsol2}\\
    R(t) = & \frac{R(0)t_0}{t_0-t}, \label{eq:PCsol3}
\end{align}
\end{subequations}
where $t_0$ is a parameter found from the initial conditions of the pinned circle.

\begin{figure}
    \centering
    \begin{subfigure}[c]{0.45\linewidth}
        \includegraphics[width=\linewidth]{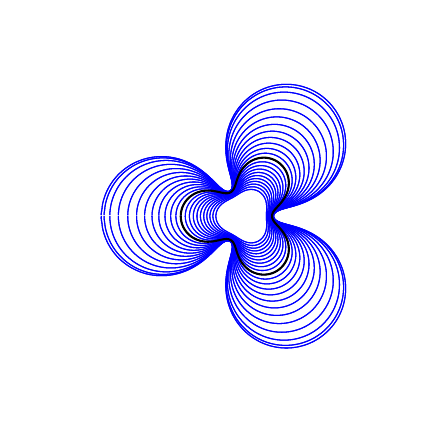}
        \caption{}
    \end{subfigure}
    \begin{subfigure}[c]{0.45\linewidth}
        \includegraphics[width=\linewidth]{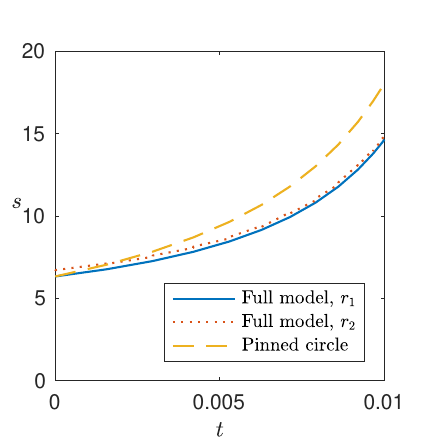}
        \caption{}\label{fig:PCradius}
    \end{subfigure}
    \caption{(a) Numerical solution of the full filament model,~\eqref{eq:FFM_vnEqn} and~\eqref{eq:FFM_hEqn}, for $\Delta P=1, \sigma=0.5$ with an $n=3$ perturbation to the centreline. (b) Semi-major and semi-minor axis (dotted red and solid blue lines, respectively) of the growing filament from the full model, compared with the predictions of the pinned circle analysis (dashed orange line,~\eqref{eq:PCsol3}). The pinned circle model is fitted at the time corresponding to the centreline marked black in (a).}
    \label{fig:PC_comparison}
\end{figure}

\begin{figure}
    \centering
    \includegraphics[width=0.55\linewidth]{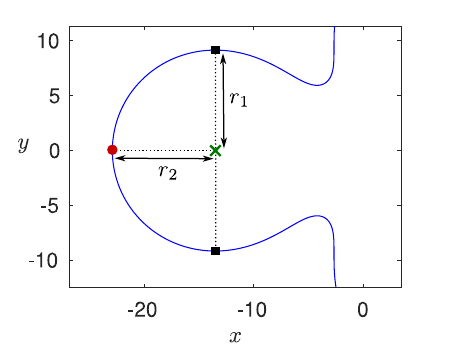}
    \caption{Extraction of the semi-minor and semi-major axes lengths, $r_1$ and $r_2$, respectively, from the identified points of maximal extent in the $x$ and $y$ directions of the circular-like filament section (blue line).}
    \label{fig:findAxes}
\end{figure}

In Fig.~\ref{fig:PC_comparison} the pinned circle asymptotic solution,~\eqref{eq:PCsol2} and~\eqref{eq:PCsol3}, is compared with the nonlinear growth of the circular-like filament sections from numerical solutions of the full filament model,~\eqref{eq:FFM_vnEqn} and~\eqref{eq:FFM_hEqn}, with an $n=3$ perturbation to the centreline. The semi-major and semi-minor axes of the filament shape are computed by finding the points of maximal extent in the $x$ and $y$ directions (marked with a red circle and black squares, respectively, in Fig.~\ref{fig:findAxes}). The distance between these and the centre of the pinned circle (marked with a green `$\times$' in Fig.~\ref{fig:findAxes}) gives $r_1$ and $r_2$ plotted in Fig.~\ref{fig:PCradius}. The initial radius, $R(0)$, of the pinned circle model is fitted at the time corresponding to the black centreline in Fig.~\ref{fig:PC_comparison}, giving a predicted blow-up time of $t_0=0.0154$ to three significant figures. The radius predicted by~\eqref{eq:PCsol3} is plotted in Fig.~\ref{fig:PCradius}.

The filament away from its pinning point is not exactly circular, as seen by the unequal semi-major and semi-minor axis lengths. These lengths converge as the filament grows. The radius predicted by~\eqref{eq:PCsol3} grows faster than the numerical solutions. Nonetheless, the qualitative behaviour of the filament's growth is captured by the pinned circle model. The numerical solution could not be run for longer time and larger radius due to the presence of numerical instabilities.

\subsection{Properties of pinned circles}

We highlight some of the properties of the pinned circles described by~\eqref{eq:PCsol1}--\eqref{eq:PCsol3}. As their name suggests, the circles are pinned at one end, i.e., the point at $\theta=\pi$ is fixed. The thickness of the filament diverges near the pinning point, and thus the model becomes inapplicable in the vicinity of $\theta=\pi$. Near these points the circular model does not describe the behaviour observed in the numerical solutions, due to the formation of `necks' connecting the pinned circles, as seen in Figs.~\ref{fig:nonlinearGrowth} and~\ref{fig:PC_comparison}.

The total mass, $m(t,\alpha)$, i.e., area, between the angle range $-\alpha<\theta<\alpha$, is
\begin{align}
    m(t,\alpha) = 2\Delta P(t_0-t)\tan\left(\frac{\alpha}{2}\right) .
\end{align}
That is, any fixed angular section is losing mass over time! The liquid is ejected toward the pinning point, located at $\theta=\pi$, which encompasses all the mass in the limit $t\to t_0$. The pinned circle solution exhibits a finite time blow-up in the radius, at the time $t_0$, in contrast to the exponential growth of the axisymmetric circles, described in Sect.~\ref{sec:filament_axisymmetric}. This is due to the mass-shedding toward the pinning point, decreasing the filament thickness and accelerating the radial growth.

\section{Conclusion} \label{sec:conclusion}

Using the newly derived filament model of Dallaston {et al.}~\cite{Dallaston2024}, the stability of fluid filaments in Hele-Shaw cells, driven by a constant pressure gradient, is studied. It is found that thin circular filaments grow if their initial radius exceeds the (dimensionless) critical radius, $R_c$ (see~\eqref{eq:sc_defn}). Further, linear stability of the axisymmetric solution reveals that all modes are stable for $R<2R_c$, with modes becoming unstable for larger radii.

A translating circular solution is found for asymptotically large radius, termed a `pinned circle'. These are thought to describe the observed non-linear growth of filament into circular-like solutions. The solutions exhibit a finite-time blow up in the pinned circle radius, attributed to the circle shedding mass as it grows. Future work will include analysing the stability of the pinned circles, and their next-order corrections in the expansion of large radius.

\bigskip

\end{document}